\documentclass{emulateapj}
\usepackage{epsfig}

\begin{document}

\slugcomment{\apj\, submitted 20 March 2011, accepted 10 July 2011}
\title{Quasisteady Configurations of Conductive Intracluster Media}
\author{G. M. Voit\altaffilmark{1}
         } 
\altaffiltext{1}{Department of Physics and Astronomy,
                 Michigan State University,
                 East Lansing, MI 48824, 
                 voit@pa.msu.edu}

\begin{abstract}
The radial distributions of temperature, density, and gas entropy among cool-core clusters tend to be quite similar, suggesting that they have entered a quasi-steady state. If that state is regulated by a combination of thermal conduction and feedback from a central AGN, then the characteristics of those radial profiles ought to contain information about the spatial distribution of AGN heat input and the relative importance of thermal conduction.  This paper addresses those topics by deriving steady-state solutions for clusters in which radiative cooling, electron thermal conduction, and thermal feedback fueled by accretion are all present, with the aim of interpreting the configurations of cool-core clusters in terms of steady-state models.  It finds that the core configurations of many cool-core clusters have entropy levels just below those of conductively balanced solutions in which magnetic fields have suppressed electron thermal conduction to $\sim 1/3$ of the full Spitzer value, suggesting that AGN feedback is triggered when conduction can no longer compensate for radiative cooling.  And even when feedback is necessary to heat the central $\sim 30$~kpc, conduction may still be the most important heating mechanism within a cluster's central $\sim 100$~kpc.
\end{abstract}

\keywords{cooling flows --- galaxies: clusters: general --- X-rays: galaxies: clusters}

\section{Introduction}

\setcounter{footnote}{0}

The cores of galaxy clusters have lessons to teach about accretion of hot intergalactic gas onto massive galaxies and the feedback processes that limit star formation within those galaxies.    Many of the most massive galaxies in the universe can be found at the centers of galaxy clusters.  A cluster's central galaxy is often the brightest cluster galaxy (BCG), and the majority of central galaxies exhibit little or no star formation.  Minimal star formation is understandable for central galaxies in which the hot intracluster gas cannot radiate its thermal energy within a Hubble time but is harder to understand in clusters with shorter central cooling times.  Star formation is generally seen in the centers of galaxy clusters only when the central cooling time is $\lesssim 1$~Gyr \citep{mo92,Cardiel+98,Rafferty+08,Bildfell+08}.  And even then, the star formation rate is usually $\lesssim 10$\% of the rate one would naively infer from the apparent cooling rate of the intracluster medium \citep{ODea+08}. 

Feedback from a central active galactic nucleus (AGN) is the favored mechanism for suppressing cooling and star formation at the centers of these clusters \citep[see][and references therein]{mn07}.  Relativistic plasma flowing from the AGN apparently evacuates large cavities within these clusters and in at least some cases drives shocks into the surrounding X-ray emitting medium \citep[e.g.][]{Forman+07,Randall+11}.  The energy introduced is comparable to that radiated from the ambient hot gas, making the hypothesis of AGN feedback seem quite plausible \citep[e.g.][]{Birzan+04}.

Yet, many questions about the AGN feedback mechanism remain unanswered.  Two of the most pressing are:  (1) how is the energy output of the AGN so well tuned to match the cooling rate of the surrounding hot gas, and (2) how does the AGN manage to distribute its energy throughout the core so that heating nearly balances cooling everywhere?  These questions arise because the AGN accretion rate would seem to depend on conditions in a neighborhood only a few tens of parsecs in size, far smaller than the $\sim 100$~kpc cluster core that must be heated.  Furthermore, the entropy gradients of cluster cores with short central cooling times tend to increase gradually from $10-100$~kpc.  A large AGN outburst that dumped most of its energy within a $\sim 10$~kpc region would overheat the center of the cluster, producing an entropy inversion \citep[e.g.][]{vd05}.  Instead, the energy source that offsets cooling somehow balances the cooling rate out to $\sim 100$~kpc without disrupting the gradually increasing radial entropy gradient.  Some implementations of AGN feedback in galaxy-cluster simulations appear promising \citep[e.g.,][]{bs09,Dubois+10,Fabjan+10,McCarthy+11}, but the problem is far from being solved.

A closer look at the data suggests that clusters with central cooling times $\lesssim 1$~Gyr do indeed achieve a quasi-steady balance between heating and cooling that sometimes allows star formation to proceed at moderate rates of $1-10 M_\odot \, {\rm yr}^{-1}$.  Gas temperatures at $\sim 10$~kpc radii in clusters with short central cooling times are 2--3 times cooler than those at  $\sim 100$~kpc, which has led such clusters to be called ``cool-core" clusters.  Gas densities decline by a factor $\sim 10$ over the same range in radius.  And even though gas density and temperature profiles in these systems differ in normalization, their profiles of entropy ($K = P \rho^{-5/3} \propto kTn_e^{-2/3}$) as a function of radius are all quite similar \citep{Donahue+06,Cavagnolo+09,sop09}.  A cluster's entropy distribution is of interest because it determines the temperature and density profiles of a cluster in hydrostatic equilibrium (see \citet{Voit+02} and \S~2).  Mergers and accretion shocks driven by gravitational structure formation produce a family of clusters with self-similar entropy profiles, and deviations from those gravitationally generated profiles reflect the action of non-gravitational cooling and heating processes \citep[e.g.][]{Voit+03}.

The purpose of this paper is to explore why one particular set of density, temperature, and entropy profiles is favored among cool-core clusters.  Its mode of exploration is to outline the properties of various families of steady-state solutions for hot gas within massive, spherically-symmetric dark-matter halos and compare them with observations of cool-core clusters.  While the outer regions of clusters are certainly not in a steady state because of ongoing accretion and merger events, the inner regions are likely to be closer to a steady state because the relevant timescales for sound waves, gravitational interactions, and radiative cooling are all $\ll 10$~Gyr at $r \ll$~100~kpc.  

Our exploration proceeds as follows.  Section~\ref{sec-heq} sets the stage by presenting some useful expressions for the structure of clusters in hydrostatic equilibrium and demonstrating that the temperature profiles of cool-core clusters are determined primarily by the shape of the underlying gravitational potential, as long as the entropy of the central gas is sufficiently low.  Section~\ref{sec-purecool} then looks at the entropy profiles expected of cluster cores in which cooling is uncompensated by feedback.  These pure-cooling profiles turn out to be nearly power-law in form, $K(r) \propto r^\alpha$ with $\alpha \sim 1.2$ at large radii.  However, the implied cooling rates of clusters in these configurations are well known to be far larger than the star-formation rates seen in BCGs.  Some form of heating must be compensating for radiative cooling, and we would like to know how the entropy profiles of clusters respond to that heating mechanism.

Section~\ref{sec-condeq} brings conduction into the mix.  Conduction in the context of cool-core clusters has a long and checkered history.  It has been proposed several times as a mechanism with the potential to balance radiative cooling in cluster cores \citep{tr83,bm86,nm01}.  However, a cluster in which electron thermal conduction balances radiative cooling is in an unstable configuration \citep{bd88,Soker03,gor08}.  Furthermore, some cluster cores have cooling rates too large to be balanced by electron thermal conduction, even if it operates at its maximum rate, unfettered by magnetic fields \citep{Voigt+02,zn03}.  The inability of conduction to suppress cooling flows has led some authors to argue that it is unimportant in cluster cores \citep[e.g.,][]{Soker10_Bfields}, and there are certainly sharp cold fronts in cluster cores across which conduction must be highly suppressed \citep{mv07}, presumably by magnetic fields running parallel to the front.  But magnetic fields cannot suppress conduction simultaneously in all directions, and recent work has shown that MHD instabilities driven by anisotropic conduction may actively sculpt the magnetic field geometry in cluster cores \citep{psl08,pq08,McCourt+10}.

This paper does not investigate the fascinating astrophysics that can result from anisotropic conduction.  Instead it adopts a phenomenological approach in which conductive heat flow runs parallel to temperature gradients, modulo a scalar suppression factor $f_c$ to account for the presence of magnetic fields.  If turbulent velocities in the intracluster medium are sufficiently large, they will overpower buoyancy-driven MHD instabilities and randomize the magnetic field, leading to a mean local suppression factor $\langle f_c \rangle \approx 1/3$ relative to a randomly chosen direction \citep{pqs10,ro10}.  We will therefore adopt $f_c = 1/3$ as a fiducial value for the scalar suppression factor.

Even though thermal conduction alone cannot stably balance radiative cooling, it remains interesting because it might be governing the triggering of AGN feedback in cluster cores.  \citet{Donahue+05} raised this possibility in an analysis of clusters with central cooling times much less than a Hubble time but with no obvious evidence for central star formation or AGN feedback.  Those clusters had core entropy levels $\sim 30-50 \, {\rm keV \, cm^2}$, corresponding to cooling times $\sim 1$~Gyr, which could in principle be offset by thermal conduction.  \citet{Cavagnolo+08,Cavagnolo+09} confirmed this relationship between core entropy and feedback in a much larger sample:  Clusters with core entropy $\lesssim 30 \, {\rm keV \, cm^2}$ often show evidence for AGN feedback while those with higher core entropy levels generally do not.  There are some clusters and groups with powerful central radio sources that appear to have core entropy levels exceeding $\sim 30 \, {\rm keV \, cm^2}$, but closer inspection reveals that these apparent exceptions have coronae of denser, lower-entropy gas in a kiloparsec-scale region surrounding the central AGN \citep{Sun09}. 
 
Several studies have shown that thermal conduction, perhaps in concert with AGN feedback, can plausibly explain the observed bimodality in core behavior \citep{Voit+08,go09,sop09,pqs10,ro10}.  Objects with core entropy $\lesssim 30 \, {\rm keV \, cm^2}$ have central cooling rates that cannot be balanced by conduction, and AGN feedback is therefore necessary to stabilize cooling.  In objects with greater core entropy, conduction is more efficient than cooling, as long as $f_c \gtrsim 0.2$, and a modest amount of intracluster turbulence may be necessary to randomize the magnetic fields and keep $f_c$ sufficiently large \citep{pqs10}.   Feedback from an AGN is not essential to counteract cooling in these objects because of the lower gas density in their cores, and both conductive and dynamical heating are capable of boosting the core entropy until the central cooling time exceeds several Gyr \citep{co08,pqs10}.  

Section~\ref{sec-condeq} addresses these issues by solving the steady-state equations for intracluster-medium configurations in which electron thermal conduction balances radiative cooling.  While conductively balanced solutions for clusters can differ greatly in density and temperature, they are restricted to a narrow locus in the $K$--$r$ plane and can be approximated with a simple power-law expression. This locus is insensitive to cluster temperature for objects warmer than $\sim 3$~keV but depends somewhat on $f_c$.  Comparing the locus of conductive balance with the entropy profiles of actual clusters shows that the entropy slopes of cool-core clusters tend to break when they cross the conductively-balanced locus.  At large radii, the $K(r)$ profiles of all clusters lie above the conductively balanced locus and have slopes consistent with gravitational structure formation.  At small radii, the entropy profiles of many cool-core clusters lie at or slightly below the critical locus for conductive balance, with slopes that track the conductively balanced configuration.  This behavior suggests that AGN feedback switches on when thermal conduction can no longer compensate for radiative cooling.

Section~\ref{sec-thermfeedback} adds the element of thermal feedback.  When conduction is present, steady-state solutions with inward accretion are not allowed unless cooling can outpace conductive heating.  This section shows how steady-state solutions including AGN heating modify the conductively-balanced configuration, allowing {\em lower} entropy levels to persist in regions where AGN heating exceeds conductive heating.  It also compares steady-state models with cooling, conduction, and thermal feedback with observations of real clusters. These comparisons show that conduction may be a more important heat source than AGN heating in the cores of many cool-core clusters and that AGN heating is usually necessary to counterbalance cooling within only the central $\sim 30$~kpc. 

Section~\ref{sec-summary} concludes by summarizing these results.

\section{Hydrostatic Equilibrium}
\label{sec-heq}

Everyone should know by now that the hot gas in a galaxy cluster is not in strict hydrostatic equilibrium.  Incompletely thermalized gas motions stimulated by mergers may provide 5\%-20\% of the pressure support \citep[e.g.,][]{lkn09}.  And heating or cooling of gas in the central region could drive either outflow or inflow.  However, because these motions are generally subsonic, hydrostatic equilibrium is likely to be an adequate approximation for the majority of relaxed-looking clusters.  This section presents some useful formulae that describe hydrostatic cluster structure and illustrate how the temperature structure of a galaxy cluster depends on the shape of its gravitational potential well and the entropy profile of the intracluster gas.  In particular, these solutions demonstrate that clusters with cool gas at small radii do not necessarily contain gas that is actively cooling.  The positive radial temperature gradients seen in cool-core clusters are primarily a reflection of the halo's gravitational potential shape and demonstrate that the potential is close to the NFW form \citep{nfw97}.

Observational limits on the gas cooling rates in galaxy clusters have established that they are not extremely large \citep{pf06}.  And as long as the cooling rate within a galaxy cluster is moderate, it will not drive the core away from hydrostatic equilibrium.  For example, the steady-state inflow velocity at radius $r$ is
\begin{equation}
  4 \, {\rm km \, s^{-1}}  \frac {\dot{M}}  {r_{\rm kpc}} 
                                \left( \frac {r \rho} {4 \times 10^{-3} \, {\rm g \, cm^{-3}}} \right) ^{-1}
\end{equation}
where $\dot{M}$ is in $M_\odot \, {\rm yr}^{-1}$, $r_{\rm kpc}$ is  $r$ in kpc, $\rho$ is the gas density, and $r \rho$ is normalized to a value typical of the inner 100 kpc of cool-core clusters \citep{vd05}.  Therefore, the inflow velocities induced by moderate rates of core cooling ($\lesssim 10 \, M_\odot \, {\rm yr}^{-1}$) should be highly subsonic in the regions resolvable with current X-ray telescopes.  

In the limit of negligible inflow velocity, the steady-state momentum equation reduces to one of hydrostatic equilibrium, which can be expressed as
\begin{equation}
 \frac {d} {dr} P^{({\gamma-1}) / { \gamma} } 
 	= - \frac {\gamma-1} {\gamma} \frac {GM_r } {K^{1/\gamma}  r^2}
  \label{eq-heq}
\end{equation}
for gas with equation of state $P = K \rho^\gamma$.  Assuming that the gaseous contribution to $M_r$ is small, one can directly integrate this equation for a given entropy profile $K(r)$, yielding
\begin{equation}
  \frac {P} {P_1} = \left[ 1 - \frac {\gamma - 1} {\gamma} 
  				\int_{r_1}^r \left( \frac {K} {K_1} \right)^{-1/\gamma}
				\frac {G M_r \mu m_p} {kT_1r} \frac {dr} {r} 
				\right] ^{\frac {\gamma} {\gamma-1}}   \label{eq-pheq}
\end{equation}
where $P_1$, $K_1$, and $T_1$ are the pressure, entropy, and temperature at some fiducial radius $r_1$.  Equation (\ref{eq-pheq}) is a generalization of the pressure solution for power-law entropy profiles given by \citet{clf09}.  Analogous hydrostatic solutions for the density and temperature profiles follow directly from the relations $\rho = (P/K)^{1/\gamma}$ and $T = (\mu m_p / k) P^{(\gamma-1)/\gamma} K^{1/\gamma}$.

These relationships can be used to clarify how a cluster's gravitational potential well and intracluster entropy gradient together determine the run of gas temperature with radius.  Realistic configurations for gravitationally confined intracluster media have substantial pressure gradients for which $P \rightarrow 0$ at large radii.  Equation (\ref{eq-pheq}) therefore implies
\begin{equation}
  T_1 \approx \frac {2 (\gamma-1)} {\gamma} \int_{r_1}^{r_{\rm b}} 
         				\left( \frac {K} {K_1} \right)^{-1/\gamma} T_\phi(r)  \; d \ln r \; \; ,
				\label{eq-t1}
\end{equation}
where $r_{\rm b}$ is a boundary radius at which $P \ll P_1$ and
\begin{equation}
  kT_\phi \equiv \frac {GM_r \mu m_p} {2r} \equiv \frac {1} {2} \mu m_p v_c^2 \; \; 
\end{equation}
defines the characteristic temperature $T_\phi$ and circular velocity $v_c$ associated with the mass $M_r$ within $r$.  In other words, equation (\ref{eq-t1}) shows that the temperature of the intracluster medium at radius $r_1$ is determined almost exclusively by the depth of the potential well and the {\em shape} of the entropy profile exterior to that radius.  It does not depend on the normalization of the entropy profile because multiplying the both the pressure and density profiles by the same factor $C$ corresponds to multiplying the entropy profile by a factor $C^{1-\gamma}$ without changing the temperature profile.  

\begin{figure*}[t]
\includegraphics[width=7.0in, trim = 1.0in 2.0in 0.7in 2.5in]{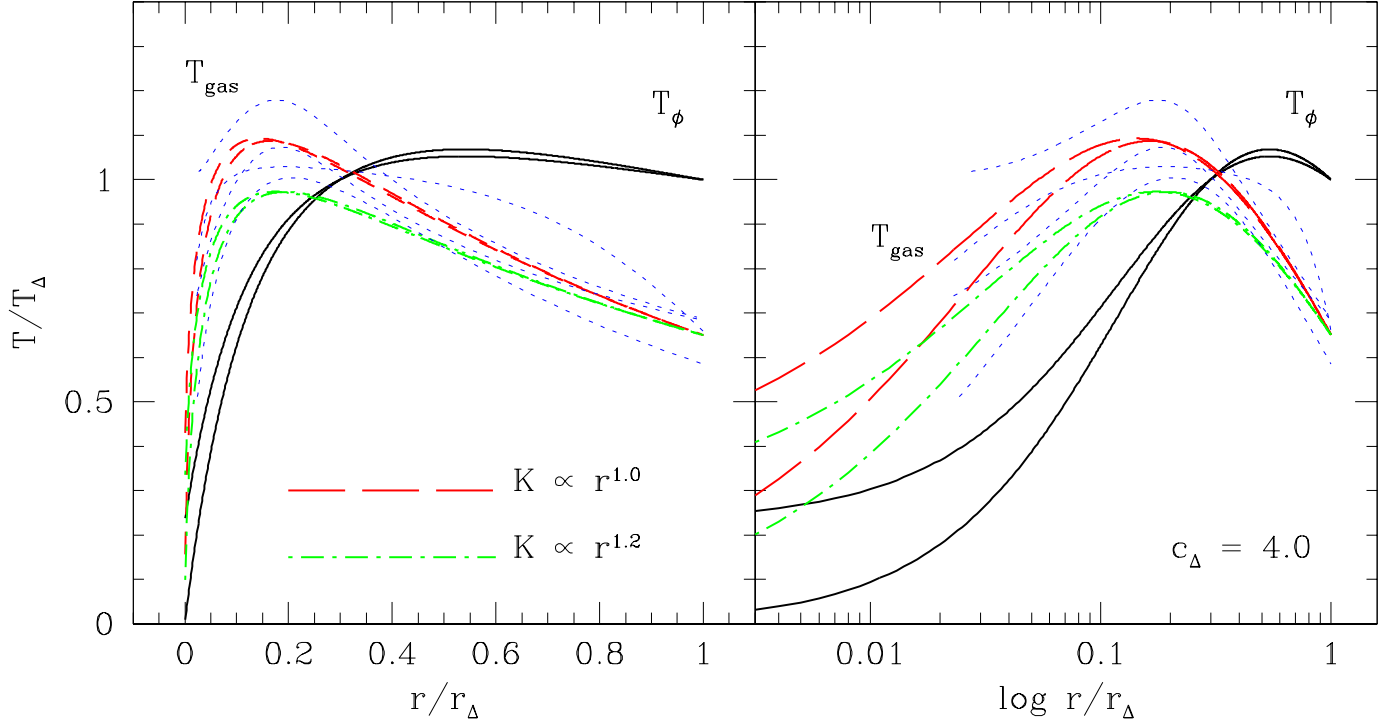} \\
\caption{ \footnotesize 
Relationship between hydrostatic gas temperature and gravitational potential for power-law entropy profiles.  Both panels show the same information, but the right-hand plot has a logarithmic radius scale to make the core structure more obvious.  Two solid lines in each panel show the characteristic gravitational temperature $T_\phi$ for halos consisting of an NFW profile with concentration $c_\Delta=4$ plus an isothermal BCG contribution amounting to $T_{\phi,{\rm BCG}}/T_{\phi,{\rm NFW}}(r_\Delta)$ = 0.0 and 0.3.  Lines that are higher at small radii correspond to a greater BCG contribution. Two dot-dashed lines show the hydrostatic temperatures for gas in those halos, assuming $K \propto r^{1.2}$ and a boundary condition $T/T_\Delta = 0.65$ at $r_\Delta$.  Long-dashed lines show the same thing for $K \propto r^{1.0}$. For comparison, dotted lines show temperature profiles (with $r_\Delta = r_{500}$) of the four clusters measured with {\em Chandra} by \citet{Vikhlinin+05,Vikhlinin+06} that have temperatures between 5~keV and 8~keV.  This comparison shows that the temperature profiles of cool-core clusters may be simply understood as resulting from intracluster gas with a nearly power-law entropy profile residing in an NFW potential.
\vspace*{1em}
\label{fig-tprofs}}
\end{figure*}

Notice also that for a typical power-law entropy profile $K \propto r^\alpha$ with $\alpha \sim 1$, the temperature at a given radius depends primarily on the value of $T_\phi$ at slightly larger radii and on the value of $\alpha$. The integral over $T_\phi$ near the boundary radius carries less weight.  For an isothermal potential, integrating over a power-law entropy profile gives $T_1 \approx [2 (\gamma-1) / \alpha ] T_\phi$, demonstrating that shallower entropy profiles lead to greater gas temperatures within a fixed gravitational potential (see also Figure~\ref{fig-tprofs}). 

The rising temperature gradients generally observed from $\sim 10$ kpc to $\sim100$ kpc in cool-core clusters therefore reflect the fact that the mass profile within $\sim 100$ kpc rises more steeply than an isothermal $M_r \propto r$ profile.  Consider the case of a gravitational potential consisting of an NFW halo with concentration $c_\Delta$ relative to the radius $r_\Delta$, for which
\begin{equation}
  T_{\phi, {\rm NFW}} (r)  \propto 
  	\frac {1} {r} \left[ \ln \left( 1+ \frac {c_\Delta r} {r_\Delta} \right) 
			- \frac {c_\Delta r/r_\Delta} {1+c_\Delta r/r_\Delta} \right]  \; \; .
\end{equation}
In order to account for the gravitational influence at small radii of a large central galaxy, one can add an isothermal mass profile corresponding to the constant value $T_{\phi, {\rm BCG}}$.  Solid lines in Figure~\ref{fig-tprofs} show the form of $T_\phi(r) = T_{\phi, {\rm NFW}}(r)  + T_{\phi , {\rm BCG}}$ for $c_\Delta = 4$ and central galaxy mass ratio $T_{\phi, {\rm BCG}} / T_{\phi, {\rm NFW}}(r_\Delta)$ equal to 0.0 and 0.3.  All of these temperature profiles are scaled to $T_\Delta \equiv T_\phi(r_\Delta)$, and in each case, $T_\phi$ steadily rises out to radii $\sim 0.3 r_\Delta$ and becomes nearly constant at larger radii.  

% ----------------------------------------
% ----------------------------------------

Temperature profiles of intracluster media within these gravitational potentials tend to peak at radii smaller than the peak of $T_\phi$, as one would anticipate from equation (\ref{eq-t1}).  Dot-dashed lines in Figure~\ref{fig-tprofs} show $T(r)$ for $\alpha = 1.2$ and the two potentials illustrated by the solid lines; long-dashed lines show the same thing for $\alpha = 1.0$.  The outer boundary condition in both cases is $T(r_\Delta)/T_\Delta = 0.65$.  As with $T_\phi (r)$, a greater BCG contribution to the potential leads to a greater temperature at small radii.  Notice that the shallower power-law entropy slope of $\alpha = 1.0$ leads to a peak temperature $\sim 20$\% greater than for $\alpha = 1.2$ and that the presence of a BCG potential prevents the inner temperature from dropping much below $\sim 0.3 \, T_\Delta$, even though the gas entropy approaches zero at small radii.  Formally, the minimum value of the gas temperature is $\sim T_{\phi,{\rm BCG}}$, and will be greater if the central entropy does not go to zero, perhaps explaining why there is little spectral evidence for gas cooler than $\sim 0.3 \, T_\Delta$ in cool-core clusters. 

Comparing these hydrostatic temperature profiles to those of real clusters shows that temperature profiles of cool-core clusters may be simply understood as resulting from intracluster gas with a nearly power-law entropy profile that resides in an NFW potential.  Dotted lines in Figure~\ref{fig-tprofs} show fits by \citet{Vikhlinin+06} to four intracluster temperature profiles of clusters between 5~keV and 8~keV observed with {\em Chandra}.  They are characteristic of other cool-core clusters and peak at radii and amplitudes similar to the peaks of the model profiles, in which the position of the gas temperature peak echoes the peak of $T_\phi$ at larger radii.  As long as $K(r) \propto r^\alpha$ with $\alpha \sim 1$, the temperature structure simply responds to the shape of the potential well.

Understanding the origin of these power-law entropy profiles is easy at large radii but less so at small radii.  Numerical simulations show that gravitational structure formation naturally produces entropy profiles with $\alpha \approx 1.1 - 1.2$ outside of cluster cores \citep{vkb05}.  There is less consensus among simulations on what $K(r)$ should be at small radii.  When radiative cooling is not allowed, adaptive-mesh hydrodynamical methods typically produce more core entropy than smooth-particle methods \citep{Frenk99_SB,vkb05}.  This discrepancy has been attributed to the greater mixing allowed in grid-based methods \citep{Mitchell+09}.  However, cooling and feedback processes strongly influence the entropy profile in cluster cores and likely dominate gravitational heating at the smallest radii \citep[e.g.,][]{Voit+03}.  Nevertheless, observations show that entropy profiles of many cool-core clusters remain close to $K \propto r^{1.2}$ at $r \ll 100$~kpc \citep{Cavagnolo+09}.  The primary goal of the rest of this paper is therefore to understand what physical processes determine the slope of a cluster's entropy profile inside the core regions, because that is what determines the radial profiles of temperature and density.

\section{Pure Cooling}
\label{sec-purecool}

Let us next consider quasisteady configurations in which pure radiative cooling determines the inner slope of the entropy profile.  Moderate cooling will cause a slow inflow, and the steady solutions we derive in this section show that the entropy slope produced by cooling at small radii is coincidentally very similar to the $K \propto r^{1.2}$ slope produced by gravitational structure formation at larger radii.

When cooling is present, we need to find solutions that satisfy both the equation of hydrostatic equilibrium and a steady-state entropy equation in which inward advection of heat balances radiative losses:
\begin{equation}
  \frac {P} {\gamma - 1} \frac {v} {r} \frac {d \ln K} {d \ln r} = - \rho^2 \Lambda(T) \; \; .
\end{equation}
Here $\Lambda(T)$ is a cooling function and we have adopted a sign convention in which negative $v$ and $\dot{M}$ correspond to mass inflow.   The physical processes determining the entropy gradient become a little clearer if we rewrite this equation as 
\begin{equation}
  \frac {d \ln K} {d \ln r} = - \frac {r} {v} \frac {(\gamma - 1) \rho^2 \Lambda} {P}  
  		= \frac {t_{\rm flow}} {t_{\rm cool}} 
\end{equation}
with 
\begin{equation}
  t_{\rm flow} \equiv - r/v   \; \; \; \;  \;  \; \; \; \;     t_{\rm cool} \equiv \frac {P} {(\gamma-1) \rho^2 \Lambda}
\end{equation}
Approximate equality between the flow time and the cooling time at all radii therefore permits an approximate power-law entropy-profile solution for a particular $\dot{M}$ determined by the normalization of the entropy profile.  

Yet another rearrangement of the steady-state entropy equation from pure cooling helps to illustrate the relationship between the entropy normalization and $\dot{M}$:
\begin{equation}
  \frac {d} {dr} K^{2/(\gamma-1)} = - \frac {8 \pi r^2} {\dot{M}} 
  					\left( \frac {kT} {\mu m_p} \right)^{\frac {3-\gamma} {\gamma-1}} 
					\Lambda (T)  \; .
	\label{eq-KgradMdot}
\end{equation}
In a pure-cooling configuration, one expects $K \rightarrow 0$ as $r \rightarrow 0$.  With this inner boundary condition and a given temperature profile $T(r)$, one can integrate to obtain
\begin{equation}
  K(r) = \left[  - \frac {8 \pi} {\dot{M}}
      		\int_0^r \left( \frac {kT} {\mu m_p} \right)^{\frac {3-\gamma} {\gamma-1}}  \Lambda (T)  r^2 dr 
					\right]^{\frac{\gamma-1} {2}}  \; .
					\label{eq-kpurecool}
\end{equation}
Notice that each pair of temperature and entropy profiles that solves equations (\ref{eq-pheq}) and (\ref{eq-kpurecool}) can be extended into a family of solutions that share the same profile shape but whose entropy normalization is $\propto \dot{M}^{(1-\gamma)/2}$.  This relationship arises because $\dot{M}$ is proportional to the square of the density scale in a pure-cooling solution, and entropy is proportional to $\rho^{1 - \gamma}$.  Furthermore, if the temperature profile is isothermal and $\gamma = 5/3$, we obtain $K \propto r$.  

\begin{figure}[t]
\includegraphics[width=3.5in, trim = 1.0in 1.4in 0.9in 1.0in]{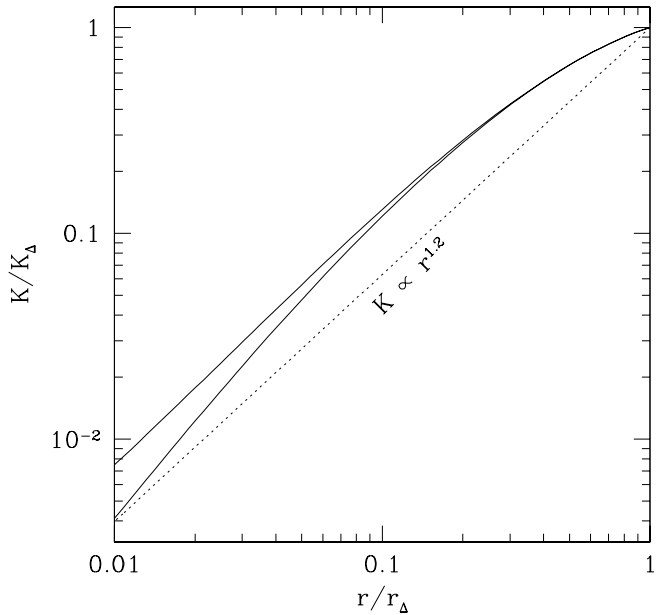} \\
\caption{ \footnotesize 
Entropy profiles for steady pure-cooling models.  Solid lines show profiles for the two gravitational potentials from Figure 1, and as in Figure 1, a greater BCG contribution leads to a higher line at small radii.  Each entropy-profile solution has boundary conditions $K/K_\Delta = 1$ and $T/T_\Delta = 0.5$ at $r = r_\Delta$, with $\dot{M}$ adjusted so that $K \rightarrow 0$ as $r \rightarrow 0$.  The dotted line shows a $K \propto r^{1.2}$ reference profile, illustrating the similarity of these steady pure-cooling solutions to that power-law profile.
\vspace*{1em}
\label{fig-kcool_profs}}
\end{figure}

Real clusters often have positive radial temperature gradients in their cores, making the effective power-law entropy slope of the pure-cooling solutions slightly steeper than linear.  Figure \ref{fig-kcool_profs} shows some steady pure-cooling solutions for the gravitational potentials in Figure~\ref{fig-tprofs} and free-free cooling with $\Lambda(T) \propto T^{1/2}$.   These solutions were obtained from equations (\ref{eq-heq}) and (\ref{eq-KgradMdot}) with the boundary conditions $K/K_\Delta = 1$ and $T/T_\Delta = 0.5$ at $r/r_\Delta = 1$, and $\dot{M}$ was adjusted so that $K \rightarrow 0$ as $r \rightarrow 0$.  They are therefore characteristic of any pure-cooling solution with realistic $K$ and $T$ at large radii and vanishing $K$ at the center.  They show that within the core of a cluster, where cooling times can be short enough for quasi-steady cooling to be a possibility, the pure-cooling solution is similar to a power law with $K \propto r^{1.2}$.  For the purpose of this plot, the entropy normalization factor $K_\Delta$ is arbitrary, because it is degenerate with the adjustable parameter $\dot{M}$. 

% ----------------------------------------
% ----------------------------------------

The agreement in power-law slope between these pure-cooling solutions and the $K \propto r^{1.2}$ power law from gravitational structure formation is purely coincidental but may help explain why the entropy profiles of cool-core clusters show no obvious break in slope where the intracluster cooling time equals the age of the universe.  It is coincidental for the following reason:  The requirement that $t_{\rm flow} \approx t_{\rm cool}$ at all radii implies $\rho \propto r^{-3/2} (T/\Lambda)^{1/2}$ and $K \propto r (T^2 \Lambda)^{1/3}$ for $\gamma = 5/3$, meaning that the index of the resulting power-law relationship between $K$ and $r$ is determined in part by the temperature dependence of the cooling function, which has nothing to do with gravitational heating.  Because of this coincidence, there is no break in slope when a structure-formation solution at larger radii ($\gtrsim 0.1 r_\Delta$) transitions to a quasisteady pure-cooling solution at smaller radii \citep[see][for simulations demonstrating this behavior]{eb08}.

The long-standing problem with such pure-cooling solutions for cluster cores is, of course, that they predict far too much cooling.  Equation (\ref{eq-KgradMdot}) can be used to show how the excessive cooling rates of pure cooling solutions are related to the normalization of the entropy profile. Solving this equation for $\dot{M}$ with $\gamma = 5/3$ and $K \propto r^\alpha$ yields
\begin{equation}
  - \dot{M} = \frac {8 \pi} {3 \alpha} \left( \frac {r} {K} \right)^{3} 
                            \left[ \left( \frac {kT} {\mu m_p} \right)^2 \Lambda(T) \right] \; \; .
                    \label{eq-mdot_K}
\end{equation}
Intracluster plasma with a temperature exceeding 2~keV, has a cooling function approximately equal to
\begin{equation}
  \Lambda(T)  =  1.7 \times 10^{-27}  \, m_p^{-2}  \, 
  				T^{1/2} \, {\rm erg \, cm^{3} \, s^{-1} \, K^{-1/2} } \; \; ,
  \label{eq-coolfunc}
\end{equation}
and the entropy profiles of clusters with cool cores tend to have $K(r) \sim 1.6 \times 10^{31} {\rm erg \, cm^2 \, g^{-5/3}} (r/1 {\rm kpc})$ \citep[e.g.,][]{Cavagnolo+09}.  In slightly less unfamiliar units, this representative entropy profile corresponds to $kTn_e^{-2/3}r^{-1} \sim 1.5 {\rm \, keV \, cm^2 \, kpc^{-1}}$.  Plugging this cooling function and entropy profile into equation (\ref{eq-mdot_K}) gives the following relationship between the normalization of the entropy gradient and the mass flow rate in the case of a steady pure-cooling solution:
\begin{eqnarray}
  - \dot{M} \; & \approx & \; 250 \, M_\odot \, {\rm yr}^{-1}   \nonumber \\
       ~ & ~ &    \; \times \left( \frac {kTn_e^{-2/3}r^{-1}} {1.5 \, \rm keV \, cm^2 \, kpc^{-1}} \right)^{-3}   
                             \left( \frac {kT} {5 \, {\rm keV}} \right)^{5/2} \; 
%                            \left( \frac {kTn_e^{-2/3}r^{-1}} {1.5 \, \rm keV \, cm^2 \, kpc^{-1}} \right)^{-3} \nonumber \\
%       ~ & ~ & \; \; \; \; \; \; \; \; \; \; \; \; \;    \times   \left( \frac {kT} {5 \, {\rm keV}} \right)^{5/2} \; \; .
  \label{eq-mdot_purecool}
\end{eqnarray}
Notice that the mass flow rate in such a steady cooling solution is highly sensitive to the normalization of the entropy profile.  Reducing the entropy by 50\% at all radii would increase the density and pressure by a factor of $2^{3/2}$ and the mass flow rate by a factor of 8.  As discussed at the end of the next section, this strong dependence of steady-state mass inflow rate on entropy level should prevent core entropy profiles from declining too far below the critical profile at which feedback starts to operate.

\section{Conductive Equilibrium}
\label{sec-condeq}

Over the years there have been many efforts to find solutions in which thermal conduction alone balances cooling, but none has been completely successful \citep[see, for example,][]{tr83,bm86,nm01}.  However, it is still useful to understand the characteristics of such solutions, because conduction in combination with AGN heating can in principle lead to quasi-steady cluster configurations \citep{rb02,gor08}, and the conductively balanced solution represents the limiting case in which AGN heating goes to zero.

In the conductively balanced case, the steady-state entropy equation that must be solved in conjunction with hydrostatic equilibrium is
\begin{equation}
  \frac {1} {r^2} \frac {d} {dr} r^2 \kappa \frac {dT} {dr} = \rho^2 \Lambda(T) \; \; ,
  \label{eq-condbal0}
\end{equation}
where the thermal conduction coefficient $\kappa$ is a function of temperature alone if electron diffusion dominates the heat flux.  Typically, the density of the intracluster medium changes with radius much more rapidly than the temperature profile.  It therefore makes sense to seek approximate solutions under the assumption that $\alpha_T \equiv d \ln T / d \ln r$ is nearly constant with radius, in which case
\begin{equation}
  \alpha_T \left[ 1 + \alpha_T (\alpha_\kappa + 1) \right]  \approx \frac {r^2} {\lambda_{\rm F}^2} \; \; ,
  \label {eq-condbal}
\end{equation}
where $\alpha_\kappa \equiv d \ln \kappa / d \ln T$ and the Field length \citep{Field65,bm90} is defined to be 
\begin{equation}
  \lambda_{\rm F} \equiv \sqrt { \frac {\kappa T} {\rho^2 \Lambda} } \; \; .
  \label{eq-kcondapprox}
\end{equation}
Cool-core clusters typically have $\alpha_T \sim 0.3$ at radii from 10--100~kpc \citep{Voigt+02,Donahue+06} and if Spitzer conduction with $\alpha_\kappa =  5/2$ applies, then the left-hand side of equation \ref{eq-condbal} is $\sim 0.6$.  The run of density with radius in a conductively balanced configuration is therefore determined by the condition that the Field length at a given radius be similar to the radius itself.

The entropy profile corresponding to equation (\ref{eq-condbal}) takes on a particularly simple form if the Field length is determined by Spitzer conduction and free-free cooling.  In that case,
\begin{equation}
  \kappa = 6 \times 10^{-7}  \, T^{5/2} \, f_c \, {\rm erg \, cm^{-1} s^{-1} K^{-7/2}} \; \; ,
\end{equation}
where $f_c$ is a scalar suppression factor depending on the magnetic field geometry. The Field length is then a function of entropy alone:
\begin{equation}
  \lambda_{\rm F} = 0.2 \, {\rm kpc} \, f_c^{1/2} \left( \frac {kTn_e^{-2/3}} {\rm keV \, cm^2} \right)^{3/2} \; \; ,
  \label{eq-lambdaf}
\end{equation}
\citep{Donahue+05}.\footnote{Astute readers may recognize that this expression for the Field length as a function of entropy is somewhat larger than in \citet{Donahue+05}. This discrepancy arises from the normalization of the cooling function.  \citet{Donahue+05} used a value of $\Lambda$ appropriate for solar-metallicity plasma at $\sim 1$~keV, which is dominated by line cooling and cannot be extrapolated to higher temperatures according to $\Lambda \propto T^{1/2}$.  This paper normalizes the cooling function so that it is appropriate for higher-temperature plasma, meaning that the assumed $T^{1/2}$ dependence leads to an underestimate of $\Lambda$, and therefore an overestimate of $\lambda_{\rm F}$, for temperatures $\sim 1$~keV.}  

\begin{figure}[t]
\includegraphics[width=3.6in, trim = 2.25in 0.9in 2.8in 0.6in]{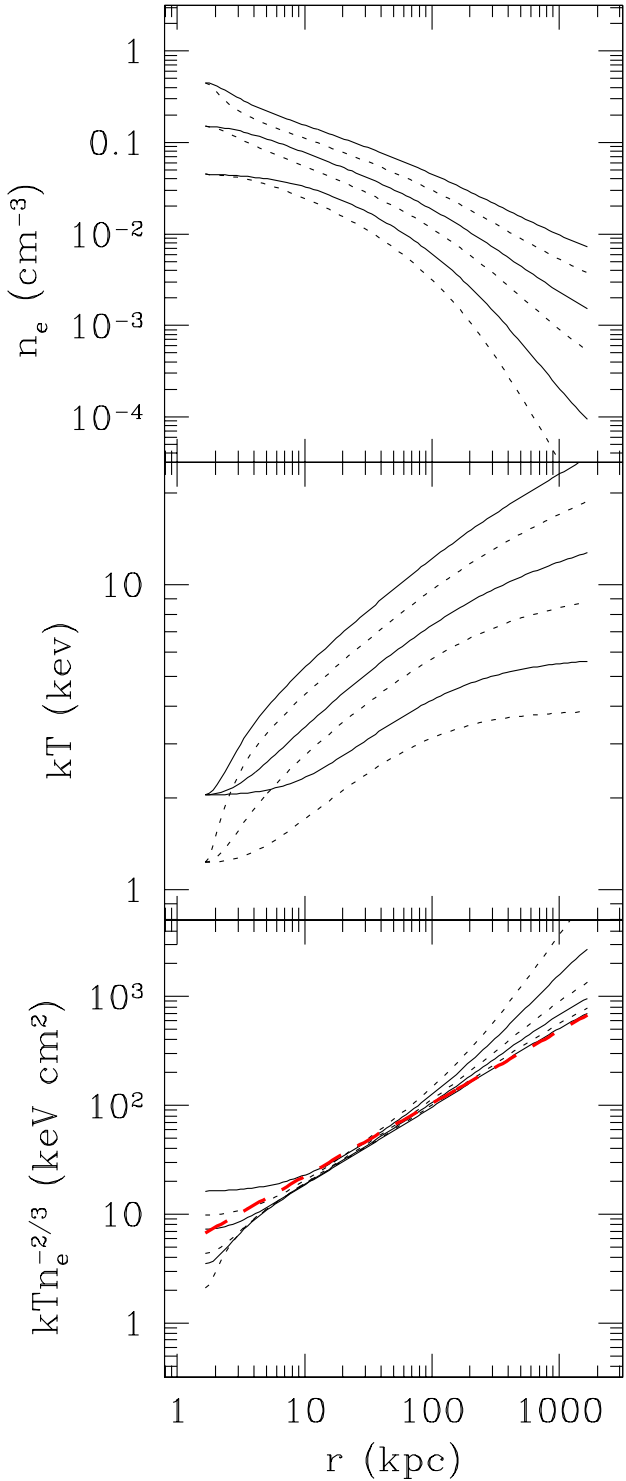} \\
\caption{ \footnotesize
Conductively balanced solutions assuming $f_c = 1/3$ for the intracluster medium of a $5 \times 10^{14} \, M_\odot$ cluster.  Each solution depends on the temperature and electron density at the inner boundary, which are $T/T_\Delta = 0.5$ (solid lines) and 0.3 (dotted lines) and $n_e =$ 0.45, 0.15, and 0.045 in cm$^{-3}$.  Solutions with greater $n_e$ profiles (upper panel) have hotter temperature profiles  (middle panel) because a greater temperature gradient is needed to offset cooling.  However, they all have similar entropy profiles (lower panel), which are close to the analytical approximation given by equation (\ref{eq-critcond_ff}) and shown by the dashed line.  
\label{fig-condbal}}
\end{figure}

Figure~\ref{fig-condbal} shows some representative solutions for conductively balanced intracluster media within the potential well of a $5 \times 10^{14} \, M_\odot$ cluster having a halo concentration of $c_\Delta = 4$.   These solutions were obtained by numerically integrating equations (\ref{eq-heq}) and (\ref{eq-condbal0}) with $f_c = 1/3$ and they depend on the values of $T$ and $n_e$ at the inner boundary.  One set of solutions (solid lines) has $kT = 0.5 kT_\Delta \approx 2 \, {\rm keV}$ and $n_e = 0.45$, 0.15, and 0.045 ${\rm cm^{-3}}$ at the inner boundary.  The other has $kT = 0.3 kT_\Delta \approx 1.2 \, {\rm keV}$ and the same set of inner density values.  For simplicity, equation (\ref{eq-coolfunc}) was used for the cooling function. 

% ----------------------------------------
% ----------------------------------------

These conductively balanced solutions have the following features.  Greater inner density leads to a more elevated temperature profile, because a steeper temperature gradient is needed to balance the additional cooling.  Greater inner temperature for a given inner density leads to a more elevated density profile, because a greater pressure is needed to confine the hotter medium.  Among these solutions, the ones with the most realistic density and temperature profiles are those with $n_e \approx 0.05 \, {\rm cm}^{-3}$ within 10~kpc, because the density profiles of solutions with larger inner densities are too shallow and represent a total amount of gas that is far too large. However, despite their differences, all of these solutions trace approximately the same locus in the $K$--$r$ plane, which is quite similar to the analytical approximation found by combining equations (\ref{eq-condbal}) and (\ref{eq-lambdaf}) with $\alpha_T = 0.3$ and $\alpha_\kappa = 5/2$:
\begin{equation}
   K(r) \approx 4.8 \, r_{\rm kpc}^{2/3} \left( \frac {f_c} {1/3} \right)^{-1/3} {\rm keV \, cm^2} \; \; .
   \label{eq-critcond_ff}
\end{equation}
This entropy profile, illustrated with a dashed line in the bottom panel of Figure~\ref{fig-condbal},  is therefore the characteristic entropy profile one expects of cool-core clusters in which electron thermal conduction balances free-free cooling.  Its power-law slope in radius is $\alpha = 2/3$, and its normalization depends on the value of $f_c$, independent of cluster temperature, as long as $\Lambda(T) \propto T^{1/2}$.  Above this line, conduction can transfer heat more quickly than radiative cooling can shed it.  Below this line, conduction cannot compete with cooling.

In groups and low-mass clusters, the characteristic entropy profile for conductive balance has a slightly greater normalization, because collisionally excited emission-line cooling increases $\Lambda(T)$ and therefore the value of $kTn_e^{-2/3}$ at which conduction can balance cooling.  Using the cooling functions of \citet{sd93}, we have derived a metallicity-dependent correction term that accounts for this additional cooling in systems with temperatures $\gtrsim 1 \, {\rm keV}$:
\begin{eqnarray}
 K(r) \; & \approx & \; 5.0 \, r_{\rm kpc}^{2/3} \left( \frac {f_c} {1/3} \right)^{-1/3} {\rm keV \, cm^2} 
 						\nonumber \\
       ~ & ~ &    \; \times \left[ 0.75 + 1.5 \, \frac {Z} {Z_\odot} \left( \frac {kT} {\rm keV} \right)^{-1.5}  
       						\right]^{1/3} .
\label{eq-critcondprof} 
\end{eqnarray}
The $Z/Z_\odot$ term in this expression is the gas metallicity in solar units.

\begin{figure}[t]
\includegraphics[width=3.5in, trim = 1.0in 1.5in 0.9in 1.0in]{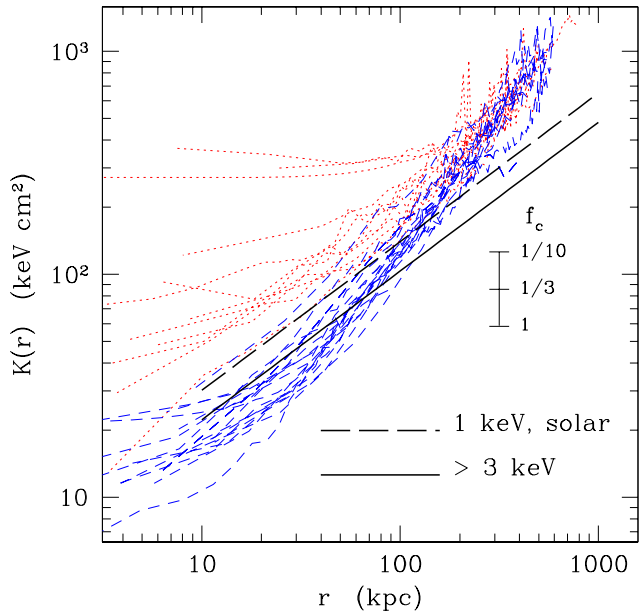} \\
\caption{ \footnotesize
Comparison of observed intracluster entropy profiles with the loci of conductively balanced profiles.  Thick lines show analytical approximations from equation (\ref{eq-critcondprof}) for conductively balanced clusters with temperatures greater than 3~keV (solid line) and for conductively balanced groups with $kT \approx 1$~keV and solar metallicity (long-dashed line).  Both lines assume $f_c = 1/3$, and the scale bar shows how the vertical positions of both lines would shift for $f_c = 1/10$ and 1.  Thin lines show observed entropy profiles for the clusters discussed in \citet{Voit+08}.  Profiles of clusters with star-forming central galaxies are illustrated with (blue) dashed lines.  Those of clusters without star formation or H$\alpha$ emission in the central galaxy are illustrated with (red) dotted lines and remain in the regime in which thermal conduction can offset radiative cooling. 
\vspace*{1em}
\label{fig-kprofcrit}}
\end{figure}

Figure~\ref{fig-kprofcrit} illustrates the approximations that equation \ref{eq-critcondprof} gives for conductively balanced profiles and compares them with the $K(r)$ profiles of actual clusters.  The thick solid line applies to clusters hotter than 3~keV, for which emission-line cooling is a small correction.  The thick dashed line shows the locus of conductively balanced profiles at lower temperature ($\sim 1$~keV) and solar metallicity.  Both lines are for $f_c = 1/3$, and a scale bar shows how the normalizations of the lines would change for $f_c = 0.1$ or 1.  Also shown are the observed entropy profiles \citep[from][]{Cavagnolo+09} of two groups of clusters from the study of \citet{Rafferty+08}.  Clusters with star-forming central galaxies (blue dashed lines) have $K(r)$ profiles that generally descend below the solid line, into the regime in which cooling operates more quickly than conduction.  Clusters with no star formation and no H$\alpha$ in the central galaxy (red dotted lines) remain within the regime in which thermal conduction is more efficient than radiative cooling, at least for $f_c \gtrsim 0.1$.  (The subset of clusters with H$\alpha$ but without a color gradient indicating star formation in the central galaxy is not shown.)

% ----------------------------------------
% ----------------------------------------

As discussed by \citet{Voit+08}, these findings suggest that the critical locus for conductive balance divides galaxy clusters into two different types.  Clusters whose entropy profiles remain entirely above this locus cannot maintain a multiphase intracluster medium because conduction will evaporate clumps of gas that are cooler and denser than the ambient medium.  With lower gas densities and longer cooling times, the cores of these clusters are also more susceptible to dynamical heating, turbulent heat transport, and mergers.  On the other hand, multiphase gas will tend to linger in the cores of clusters whose entropy profiles drop below the conductive locus, allowing them to harbor emission-line nebulae and ongoing star formation \citep[see also][]{nb04}.  Because of their low central entropy levels, these clusters have cool, dense cores that are more difficult to disrupt.

Conductive balance in this context should not be confused with thermal stability.  Conductively balanced solutions for cluster cores are well known to be unstable \citep[e.g.][]{bd88,Soker03,gor08}.  Conversely, environments that are out of conductive balance are not necessarily prone to local thermal instabilities and growth of multiphase structure.  This means that the critical locus in Figure~\ref{fig-kprofcrit} is more meaningful as an indicator of the need for thermal feedback, in addition to the heat supplied by conduction, than as a threshold for local thermal instability.   

A closer look at the entropy profiles in Figure~\ref{fig-kprofcrit} supports the idea that the locus of conductive balance is a critical threshold for triggering thermal feedback.  Above that locus, the entropy profiles shown by the dashed lines have $\alpha \sim 1 - 1.2$ slopes expected of both gravitational heating and pure cooling.  But below that locus these entropy profiles become shallower, with slopes that resemble the $\alpha = 2/3$ slope of the locus itself.  One way to understand this behavior is to consider what happens to a cluster when conduction can no longer balance core cooling.  As cooling proceeds, entropy levels in the core will drop and the inward mass flux will rapidly increase from zero toward the prodigious pure-cooling rate of equation (\ref{eq-mdot_purecool}).  If thermal feedback does indeed depend on accretion of intracluster gas at the cluster's center, then it should achieve a quasi-steady balance with radiative cooling with an entropy profile not far below the locus of conductive balance.  This may account for the similarity in core entropy, density, and temperature profiles among cool-core clusters.

\section{Thermal Feedback}
\label{sec-thermfeedback}

Most of the recent attempts to explain how cluster cores stave off catastrophic cooling have focused on feedback from an active galactic nucleus in the central galaxy.  The heating per unit volume caused by feedback can be expressed with a term ${\cal H}$ that depends on the mass accretion rate at the center of the cluster.  If conduction is also operating, the steady-state entropy equation then becomes
\begin{equation}
  \frac {P} {\gamma - 1} \frac {v} {r} \frac {d \ln K} {d \ln r} = {\cal H} - \rho^2 \Lambda
  				+ \frac {1} {r^2} \frac {d} {dr} r^2 \kappa \frac {dT} {dr}
		\label{eq-thermal_fb}
\end{equation}
The right hand side of this equation must be negative everywhere in order to obtain a solution with steady inflow ($v < 0$) and a positive entropy gradient.  Even if the accreting matter condenses into a cold phase before feeding the active nucleus, equation (\ref{eq-thermal_fb}) will remain valid outside the region where the condensation rate is a significant fraction of the mass inflow rate.

This section uses equation (\ref{eq-thermal_fb}) to explore steady-state core configurations in which conduction, radiative cooling, and thermal feedback are all significant.  It first presents a simple formalism for expressing the radial distribution of thermal feedback and its dependence on the mass accretion rate.  It then attempts to infer the radial distribution of thermal feedback from observations of actual clusters, under the assumption that they are in a steady state with $f_c = 1/3$.  The main findings of this section are that thermal conduction may often be the most important source of heat in a cluster's core and that thermal feedback, when it is necessary, is rarely required beyond $\sim 30$~kpc from the cluster's center.

\subsection{A Simplistic Model for AGN Feedback}

Thermal feedback from a central AGN can be phenomenologically modeled by expressing the outward energy flux available for heating as
\begin{equation}
  F_{\rm AGN} = \frac {\epsilon (-\dot{M}) c^2} {4 \pi r^2}  h(r) 
\end{equation}
where $\epsilon$ is an efficiency factor for thermal feedback and $h(r)$ is a function describing how that energy flux declines with radius as it is converted into heat energy within the intracluster gas.  The corresponding volumetric heating rate is 
\begin{equation}
   {\cal H}(r) =  \frac {\epsilon \dot{M} c^2} {4 \pi r^2}  \frac {dh} {dr} \; \; .
\end{equation}
According to the sign convention adopted here, both $\dot{M}$ and $dh/dr$ are negative when AGN feedback is operating.  

Grouping the terms proportional to $\dot{M}$ facilitates a direct comparison of heat advection with thermal feedback:
\begin{eqnarray}
 \lefteqn{ \hspace*{-1.5cm}
       \frac { \dot{M} c^2} {4 \pi r^3}  \left[ \frac {kT} {(\gamma - 1) \mu m_p c^2} \frac {d \ln K} {d \ln r} 
  	- \epsilon h \frac {d \ln h} {d \ln r} \right]  = }   \nonumber \\
  &  &   \hspace{2.0cm} - \, \rho^2 \Lambda \, + \, \frac {1} {r^2} \frac {d} {dr} r^2 \kappa \frac {dT} {dr}
\end{eqnarray}
The quantity in square brackets is always positive for clusters with a rising entropy gradient, and the relative importance of the two terms within the brackets depends on how the thermal energy of the gas compares with its rest-mass energy times the efficiency factor $\epsilon$.  For a typical ICM temperature of $\sim 5$~keV, the magnitude of the advection term within the square brackets is $\sim kT/\mu m_p c^2 \sim 10^{-5}$.  A value of $\epsilon$ significantly exceeding $10^{-5}$ is therefore necessary for feedback to suppress cooling and inflow.  When this is the case, the mass-inflow rate needed to maintain a steady state is inversely proportional to $\epsilon$, and as long as the configuration is close to hydrostatic equilibrium, steady-state solutions with identical values of $\epsilon \dot{M}$ will be indistinguishable.  Furthermore, the presence of a non-negligible heating term in the steady-state solution allows the entropy at radii where heating is significant to be smaller than in a conductively balanced solution. 

\subsection{Inferring the Radial Distribution of Thermal Feedback}

The responsiveness of steady-state entropy profiles to the radial distribution of heat input raises the possibility that one might be able to infer the radial distribution of feedback heating in real clusters from observations of their temperature and density profiles.  If a cluster's core configuration is truly steady, the observed density and temperature profiles can be plugged into the following equation to determine the steady-state distribution of heat input:
\begin{equation}
  \frac {dh} {d r} = \frac {4 \pi r^2} {\epsilon \dot{M} c^2} 
  	\left[ \rho^2 \Lambda -  \frac {1} {r^2} \frac {d} {dr} r^2 \kappa \frac {dT} {dr} \right] \; \; .
\end{equation}
Unfortunately, cluster temperature profiles are often too noisy for differentiating twice, so instead we will take advantage of the hydrostatic model of \S~\ref{sec-heq}.  It gives temperature profiles in terms of a gravitational potential represented by $T_\phi (r)$ and an entropy profile shape.  In what follows, the potential is assumed to be a simple NFW potential parametrized by $T_\Delta$ and $c_\Delta$, and the entropy profile is parametrized using the fitting formula of \citet{Cavagnolo+09}:  $K(r) = K_0 + K_{100} (r/100 \, {\rm kpc})^\alpha$.  For most of the clusters considered here, this five-parameter hydrostatic model adequately describes the observed temperature and density profiles.

% ----------------------------------------
\begin{figure}[t]
\includegraphics[width=3.62in, trim = 2.25in 0.9in 2.8in 0.6in]{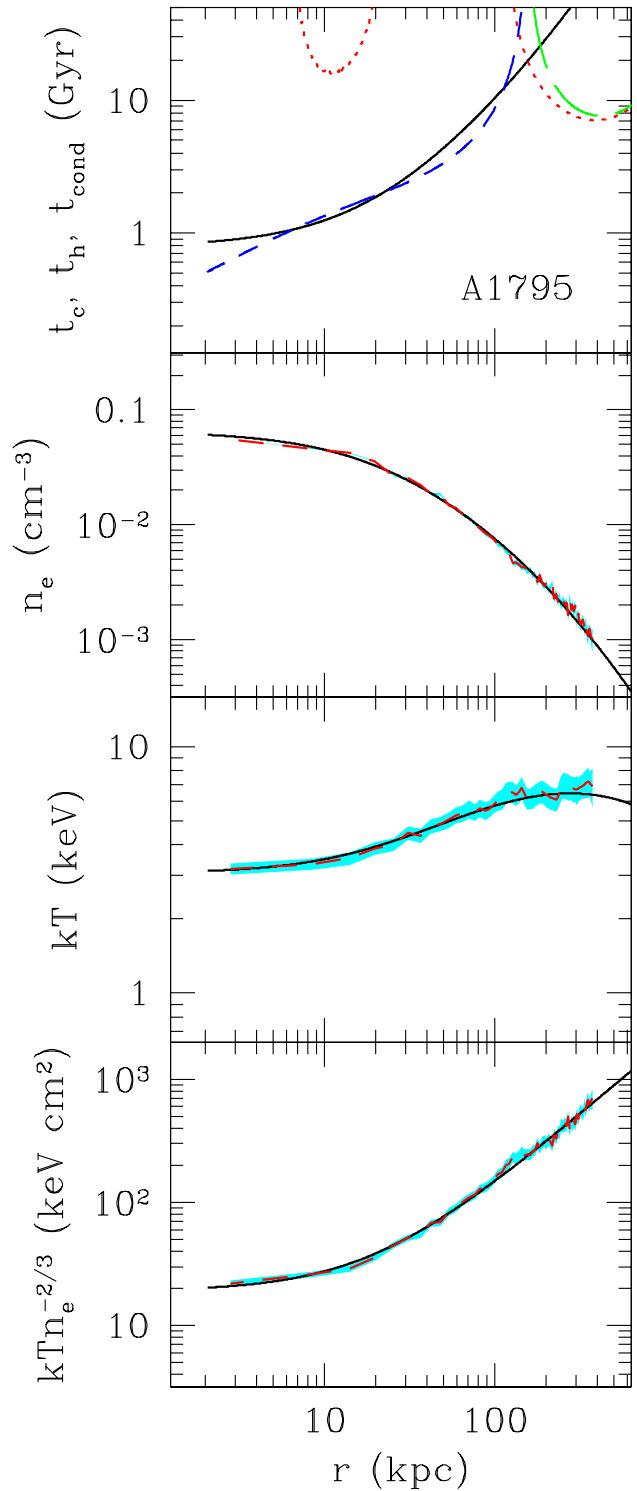} \\
\caption{ \footnotesize
A quasi-steady solution for Abell~1795.  The top panel shows the relative importance of cooling (solid line), conduction (long-dashed lines), and thermal feedback (dotted line) in terms of timescales.  The lower a line is in this panel, the more rapid and important the process.  Solid lines in the lower three panels show the density, temperature, and entropy profiles, respectively, of the solution used to compute the cooling and conduction rates from which heating was inferred.  Dashed lines in those panels show the observations and shading shows uncertainty ranges for these quantities from the database of \citet{Cavagnolo+09}.
}
\label{fig-heat_infer_A1795}
\end{figure}
% ----------------------------------------

Figure~\ref{fig-heat_infer_A1795} shows an example of a cluster that may be close to conductive balance.  The bottom three panels show the model used to infer the radial distribution of thermal feedback in Abell~1795 and how it compares with the observed density, temperature, and entropy profiles of this cluster from the compilation of \citet{Cavagnolo+09}.  The top panel shows the relative importance of heating, cooling, and conduction as a function of radius, assuming that this cluster is in a quasisteady state with  negligible advection of heat.  Under this assumption, the heating rate can be inferred from equation~(\ref{eq-thermal_fb}) by setting the left-hand side to zero and plugging in the model profiles of temperature and density.  A suppression factor $f_c = 1/3$ has been assumed.  A solid line in the top panel shows the cooling time $t_c$, defined as the quotient of the heat content at constant pressure, $5 \rho kT / 2 \mu m_p$, and the radiative cooling rate.  The dashed lines (blue and green) show the conduction time $t_{\rm cond}$, equal to the quotient of the heat content and the conduction term in equation~(\ref{eq-thermal_fb}).  The dotted line (red) shows the inferred heating time $t_h$ determined by dividing the heat content by the difference of the cooling and conduction terms.

Within 100~kpc, the timescales for cooling and conduction in Abell~1795 are nearly equal, meaning that the conduction and cooling terms in equation~(\ref{eq-thermal_fb}) nearly offset each other \citep[see also][]{zn03}.  Thermal feedback is therefore not necessary to keep this system in a steady state, and the inferred heating term corresponds to a long timescale ($> 10$~Gyr).  Outside of 100~kpc, where the temperature profile peaks, conduction acts as a coolant channeling heat away from the peak.  The line representing $t_{\rm cond}$ is green in this region instead of blue to indicate the change in sign of the conduction term.  Under the steady-state assumption, heating must compensate for conductive cooling here, and the red dotted line for $t_h$ indicates the implied heating rate.  However, the timescales in this region are comparable to the age of the universe, so the steady-state assumption is unlikely to be valid.  In fact, mergers and other forms of dynamical heating not included in equation~(\ref{eq-thermal_fb}) are relevant on these timescales.  We will therefore restrict our attention to radii $< 100$~kpc, where the steady-state assumption has a greater chance of being applicable.

% ----------------------------------------
\begin{figure}[t]
\includegraphics[width=3.62in, trim = 2.25in 0.9in 2.8in 0.6in]{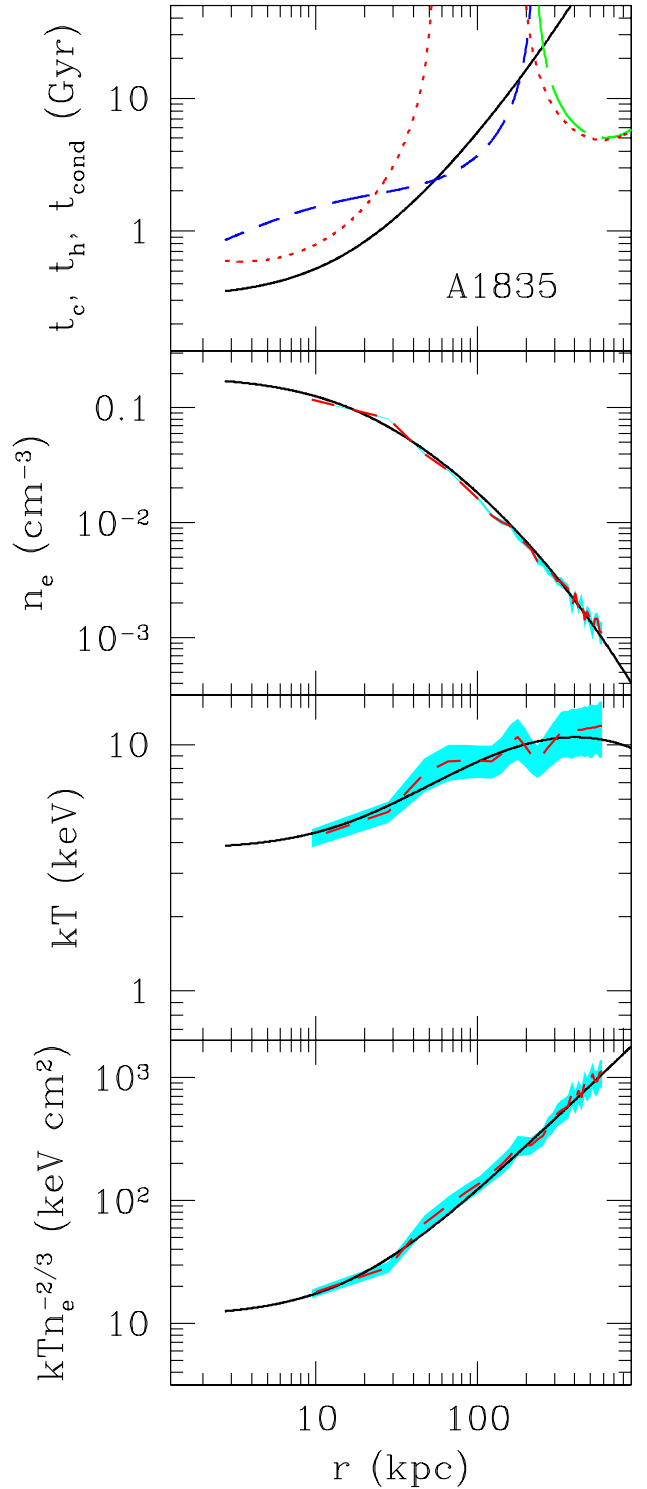} \\
\caption{ \footnotesize
A quasi-steady solution for Abell~1835.  The top panel shows the relative importance of cooling (solid line), conduction (long-dashed lines), and thermal feedback (dotted line) in terms of timescales.  The lower a line is in this panel, the more rapid and important the process.  Solid lines in the lower three panels show the density, temperature, and entropy profiles, respectively, of the solution used to compute the cooling and conduction rates from which heating was inferred.  Dashed lines in those panels show the observations and shading shows uncertainty ranges for these quantities from the database of \citet{Cavagnolo+09}.
}
\label{fig-heat_infer_A1835}
\end{figure}
% ----------------------------------------

Figure~\ref{fig-heat_infer_A1835} shows a different example, the cluster Abell~1835.  The lines represent the same things as in Figure~\ref{fig-heat_infer_A1795}, but the implications are different.  Thermal conduction in this system cannot offset radiative cooling at small radii.  The inferred thermal feedback rate within that region is greater than the conductive heating rate, causing the heating timescale $t_h$ shown in the figure (dotted line) to be shorter than the conductive timescale $t_{\rm cond}$ (dashed line), but only within $\lesssim 30$~kpc of the cluster center.

\begin{figure}[t]
\includegraphics[width=3.5in, trim = 1.0in 1.5in 0.9in 1.0in]{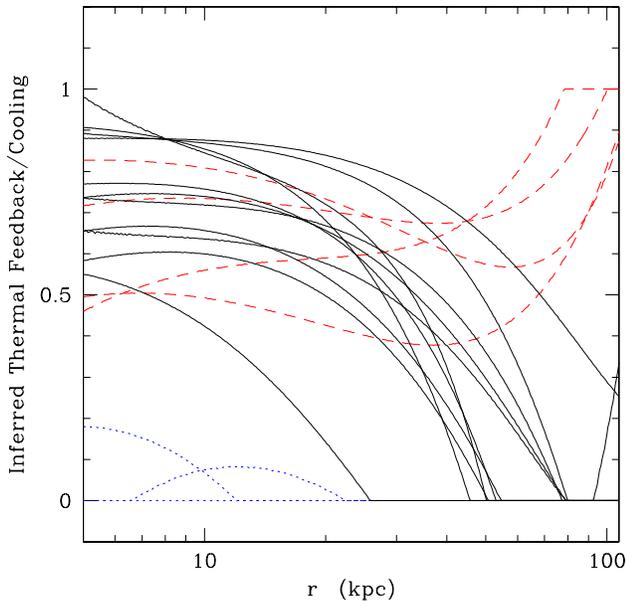} \\
\caption{ \footnotesize
Inferred ratio of thermal feedback to radiative cooling as a function of radius for conductive clusters ($f_c = 1/3$) assumed to be in a steady state.  These clusters are the 18 cool-core clusters from \citet{Rafferty+08} with star-forming central galaxies.  Dotted (blue) lines show the three clusters that are nearly in conductive balance.  Solid lines show the 11 clusters in which conduction is more important than thermal feedback outside of the central $\sim 40$~kpc.  Dashed (red) lines show the four clusters in which thermal feedback is needed to maintain a steady-state in the outer parts of the cluster core.
\vspace*{1em}
\label{fig-heatrats}}
\end{figure}

In that respect, Abell~1835 is representative of most of the clusters with star-forming central galaxies in the sample of \citet{Rafferty+08}.  Figure~\ref{fig-heatrats} shows the ratio of inferred heating to cooling as a function of radius in all 18 of those clusters.  Only three of those systems (dotted lines) are close to conductive balance at all radii.  Another 11 (solid lines) require thermal feedback in order to remain in a steady state, at least for $f_c = 1/3$, but in most of those cases inferred feedback heating dominates over conductive heating only within $\lesssim 30$~kpc.  There are, however, four systems in which the inferred steady-state heating rate remains large out to 100~kpc.  Part of the reason for the rising level of implied heating near 100~kpc in these systems is that the temperature profile is reaching its maximum value there, causing a decline in the conduction rate inferred from the temperature gradient.  It is possible that other forms of diffusive heat transport, such as turbulent transport, could be more important than conduction in these regions, reducing the need for thermal feedback there.

% ----------------------------------------
% ----------------------------------------

\begin{figure}[t]
\includegraphics[width=3.5in, trim = 1.0in 1.5in 0.9in 1.0in]{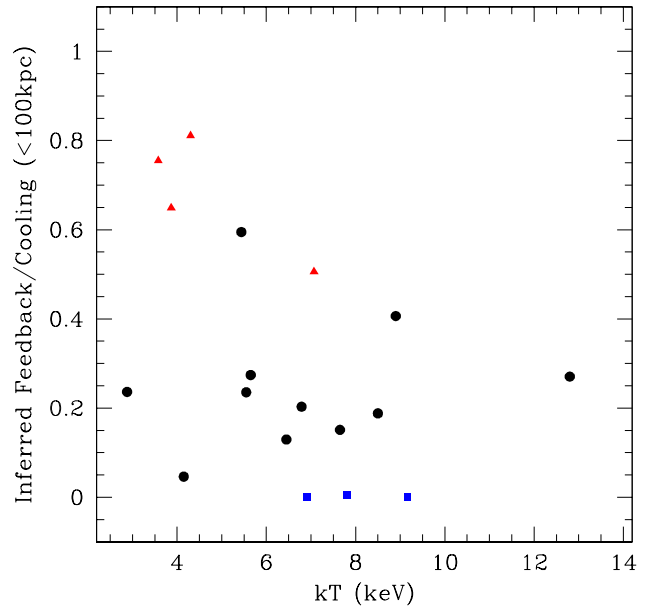} \\
\caption{ \footnotesize
Inferred ratio of thermal feedback to radiative cooling integrated within 100~kpc for the same clusters as in Figure~\ref{fig-heatrats}.   Squares show clusters close to conductive balance.  Circles show  clusters in which conduction is more important than thermal feedback outside of the central $\sim 40$~kpc.  Triangles show the four clusters in which thermal feedback is needed to maintain a steady-state in the outer parts of the cluster core.
\vspace*{1em}
\label{fig-heatrats_tot}}
\end{figure}

Integrated over the whole cluster core, conduction may be a more important heat source than thermal feedback.  Figure~\ref{fig-heatrats_tot} compares the volume integral of the implied heating rate within a radius of 100~kpc to the integrated cooling rate.  In most cases the inferred amount of thermal feedback is less than 50\% of the total cooling, meaning that conduction with $f_c = 1/3$ can replace the majority of the heat losses in these systems.  There does, however, appear to be a trend in the upper envelope of this relationship.  At lower temperatures there are more objects in which thermal feedback seems to be the dominant heat source.  It would appear that conduction is always important in the highest-temperature cool-core systems but is less effective at compensating for cooling in at least some of the lower-temperature cool-core systems. 

% ----------------------------------------
% ----------------------------------------

Taken as a whole, these results are in alignment with the findings of \citet{Voigt+02}, \citet{zn03}, and \citet{vf04}.  Those studies also found that thermal conduction could plausibly balance radiative cooling in some cluster cores, but that it fails to balance cooling within the central $\sim 30$~kpc in many others, necessitating a feedback response to prevent cooling rates in excess of the observed limits.  This study differs from those earlier efforts in that it fits a physically motivated hydrostatic model to the cluster data in order to obtain a numerically differentiable temperature profile, which it then uses to infer the radial distribution of thermal feedback.  Those distributions of heat energy are unlikely to be accurate in detail, because cool-core clusters are probably not in a perfectly steady state, and because isotropic conduction with $f_c = 1/3$ is probably overly simplistic, but they do indicate that AGN feedback does not necessarily need to match radiative cooling all the way out to $\sim 100$~kpc.  Heat input in the vicinity of the observed X-ray cavities may be all that is necessary.

\vspace*{1em}

\section{Summary}
\label{sec-summary}

Inspired by the fact that the radial profiles of intracluster gas density, temperature and entropy in cool-core clusters bear a strong resemblance to one another, this paper has tried to identify the physical processes that govern the form of those profiles.  Because the similarity among these profiles extends to within $\sim 10$~kpc of the cluster's center, where cooling, heating, and dynamical processes can operate on timescales $\lesssim 100$~Myr, we have assumed that cool-core clusters have settled into a quasi-steady state that is close to hydrostatic equilibrium.   Proceeding from that assumption, the paper derived the general characteristics of various quasi-steady configurations for the intracluster medium.

First, it showed in \S~2 that the temperature profiles of cool-core clusters simply reflect the shape of the underlying gravitational potential, as long as the intracluster entropy profile has a moderately positive radial gradient.  When that is the case, the gas temperature at each radius is determined primarily by the value of $T_\phi \propto M(<r)/r$ at slightly larger radii and to a lesser degree by the slope of the entropy profiles.  The temperature does {\em not} depend on the absolute value of gas entropy---only the slope of the entropy profile matters.  For an NFW-like potential with a concentration $c_\Delta \sim 4$, the gas temperature peaks in the $0.1 r_\Delta$---$0.2 r_\Delta$ range, and within a given potential well, configurations with shallower entropy slopes peak at greater temperatures and smaller radii.  Adding a small isothermal potential at the center to represent the mass of a central cluster galaxy helps to explain why clusters contain so little gas at temperatures less than 30\% of the virial temperature.

Next, \S~3 derived a set of pure-cooling solutions in which inward advection of heat balances radiative cooling.  Steady cooling of gas in a cluster potential turns out to produce entropy profiles similar in slope to the $K(r) \propto r^{1.2}$ profile generated by gravitational structure formation.  This coincidence accounts for why there is no break in slope in the entropy profiles of cool-core clusters at radii where the cooling time is similar to the age of the universe.  Furthermore, the mass-inflow rates of these steady pure-cooling solutions are $\propto K^{-3}$ and greatly exceed the observed star-formation rates of central galaxies in cool-core clusters.  This implies that entropy profiles in cool-core clusters should never drop far below the critical profile at which inflow begins to fuel feedback.  Once inflow begins, the feedback response should rapidly rise as the entropy level at a given radius drops.

Section~4 derived the characteristics of solutions that are in conductive balance, showing that despite widely differing density and temperature profiles, steady configurations with the same value of $f_c$ trace the same locus in the $K(r)$--$r$ plane.  Simple expressions for this locus are given for hot systems ($> 3$~keV) and for cooler systems in which the cooling rate is metallicity dependent.  Comparing the observed entropy profiles of galaxy clusters to these loci shows that clusters without star-forming central galaxies or other evidence of multiphase gas all lie above the locus for conductive balance, suggesting that conduction is preventing radiative cooling from producing a multiphase intracluster medium that can accrete into the central galaxy \citep[see also][]{Voit+08}.  On the other hand, the entropy profiles of classic cool-core clusters tend to drop below this critical locus at small radii, but they do not drop very far below it.  Instead, their entropy slopes bend in the vicinity of the critical locus and tend to track it at slightly lower entropy levels, suggesting that AGN feedback is triggered when an entropy profile drops below the critical locus and prevents the entropy from dropping much further. This would explain why the density, temperature, and entropy profiles of cool-core clusters are so similar to one another.

Section~5 considered cluster configurations in which cooling, conduction, and thermal feedback all have roles to play in maintaining a steady state.  Instead of assuming a particular model for thermal feedback, the paper attempted to infer the spatial distribution of feedback heating from observations of cool-core clusters.  These were fit to a five-parameter model in which the gravitational potential has an NFW form and the intracluster entropy profile is a power law plus a constant.  This model provides a smooth temperature profile that can be differentiated to obtain the heat deposited by thermal conduction as a function of radius.  If heat advection is assumed to be negligible, the steady-state heating rate is then equal to the difference between radiative cooling and conductive heating.  Some conductive cool-core clusters do not require thermal feedback to offset cooling, but most of them do.  However, the inferred thermal feedback rates usually exceed conductive heating only within $\lesssim 30$~kpc, where X-ray cavities are most often found.  Conduction with $f_c = 1/3$ can account for most of the heating required beyond that radius, and when integrated over the entire core out to 100~kpc, conduction is a more important heat source than thermal feedback in most of the clusters considered here.  If conduction is indeed as important as these results imply, then it is essential to include it in numerical simulations that seek to accurately model the core structures of galaxy clusters, including the stellar content of the brightest cluster galaxy.

\vspace*{1.0em}

The author thanks Peng Oh, Ken Cavagnolo, and Marcus Br\"uggen for helpful comments on the manuscript.  This work was partially supported at MSU by NSF grant AST-0908819 and {\em Chandra} Theory program grant TM9-0008X.  It was completed at the Kavli Institute for Theoretical Physics in Santa Barbara, supported in part by the National Science Foundation under Grant No. NSF PHY05-51164.

%\bibliography{conduction}

\end{document}